\documentstyle[amssymb,12pt]{article}

\begin{document}

\pagestyle{myheadings}
\markright{Li{\' e}nard-Wiechert Potentials in Even Dimensions}

\title{Li{\' e}nard-Wiechert Potentials in Even Dimensions}

\author{Metin G{\" u}rses$^{1}$\thanks{email: gurses@fen.bilkent.edu.tr} ~~and {\" O}zg{\" u}r Sar{\i}o\u{g}lu$^{2}$\thanks{email: sarioglu@metu.edu.tr}}

\begin{titlepage}
\maketitle
\begin{center}
$^1${\small Department of Mathematics, Faculty of Sciences,\\
Bilkent University, 06800, Ankara - Turkey\\}
$^2${\small Department of Physics, Faculty of Arts and  Sciences,\\
Middle East Technical University, 06531, Ankara - Turkey}
\end{center}

\begin{abstract}
The motion of point charged  particles is considered in an even dimensional
Minkowski space-time. The potential functions corresponding to the massless
scalar and the Maxwell fields are derived algorithmically. It is
shown that in all even  dimensions particles lose energy due
to  acceleration.
\end{abstract}

\end{titlepage}

\section{Introduction}

Recently Gal'tsov \cite{gal} and Kazinski et al \cite{kaz} have considered
the Lorentz-Dirac equation for a radiating point charge in a Minkowski
space-time of arbitrary dimension. They showed that the mass
renormalization is possible only in three and four dimensions. In their discussion, they have also given the retarded Green's functions of the D'Alembert equation in any dimensions which was in fact constructed rigorously a long time ago \cite{chil}.
Motivated by these works, we are interested in the radiation problem
of accelerated point charges in all even dimensions
\footnote{For the reason why we did not consider odd dimensions, please see
Appendix B.}.
Here we find the Li{\' e}nard-Wiechert potentials corresponding to the
massless scalar and the Maxwell fields in all even dimensions. We then use
these potentials to relate the radiation from an accelerated point particle
to its motion and the geometry of its trajectory. We derive the energy flux
for this radiation and show that accelerating point charged particles lose
energy in all even dimensions.

\noindent

In the next Section, we develop the kinematics of a curve $C$ in a $D$-
dimensional Minkowski manifold ${\bf M_{D}}$. In Section 3 we find the
Li{\' e}nard-Wiechert potentials of massless free scalar fields in an even
dimensional Minkowski space. We calculate the energy radiated due to the
acceleration. We show that in all even dimensions such particles lose
energy, as can be expected. In Section 4, we determine the
Li{\' e}nard-Wiechert potentials for the Maxwell theory. We give 
a recursion relation between the vector potentials
of the theory in two consecutive even dimensions. In Section 4, we also show
that particles carrying electric charges lose energy in all even dimensions.
We construct explicit solutions of the
electromagnetic vector field due to the acceleration of charged particles
in 4,6,8,10 dimensions.
We then find the energy fluxes in 4,6,8 dimensions due to  acceleration.
In  Appendix A, we give the Serret-Frenet equations in an arbitrary
Minkowski
space-time and  also some auxiliary tools used in the calculation
of the energy flux integrals. In  Appendix B, we give a proof of the
recursion relation introduced in Section 4.

\section{Curves in D-Dimensional Minkowski Space}

In our previous works \cite{gu1}, \cite{ozg}, \cite{gu2}
we have developed a curve kinematics to be utilized in finding new solutions
and in calculating energy fluxes due to the acceleration in the framework of
Einstein's general theory of relativity. Here we use the same approach
to solve the scalar and Maxwell field equations in all even dimensions.
For this purpose, we shall now give a summary of the geometry of a regular
curve in ${\bf M_{D}}$, Minkowski space-time manifold of dimension $D$.

Let $z^{\mu}(\tau)$ describe a smooth curve $C$ in ${\bf M_{D}}$,
where $\tau$ is the arclength parameter of the curve. From an arbitrary
point $x^{\mu}$ outside the curve, there are two null lines intersecting the
curve $C$. These points are called the retarded and the advanced times. Let
$\Phi$ be the distance (world function) between the points
$x^{\mu}$ and $z^{\mu}(\tau)$, then by definition it is given  by

\begin{equation}
\Phi={1 \over 2}\,\eta_{\mu \nu}\,(x^{\mu}-z^{\mu}(\tau))\,(x^{\nu}-z^{\nu}(\tau)),
\label{dist}
\end{equation}

\noindent
where $\eta_{\mu \nu}= \mbox{diag} (-1,1,\cdots,1)$.
Hence $\Phi$ vanishes at the retarded, $\tau_{0}$, and advanced,
$\tau_{1}$, times. In this work we shall
focus ourselves to the retarded case only.
The Green's function for the vector potential chooses this point on the
curve $C$ \cite{brt}, \cite{jack}.
By differentiating $\Phi$ with respect to $x^{\mu}$ and letting
$\tau=\tau_{0}$, we get

\begin{equation}
\lambda_{\mu} \equiv \tau_{,\mu}={x_{\mu}-z_{\mu}(\tau_{0}) \over R}, ~~~
R \equiv  \dot{z}^{\mu}(\tau_{0})\,(x_{\mu}-z_{\mu}(\tau_{0})),
\label{lam}
\end{equation}

\noindent where $R$ is the retarded distance, $\lambda_{\mu}$ is a
null vector and a dot over a letter denotes differentiation with
respect to $\tau_{0}$. The derivatives  of $R$ and
$\lambda_{\mu}$, using (\ref{lam}),  are given by

\begin{eqnarray}
\lambda_{\mu, \nu}&=&{1 \over R}\,[\eta_{\mu \nu}-\dot{z}_{\mu}\, \lambda_{\nu}
-\dot{z}_{\nu}\, \lambda_{\mu}-(A-\epsilon)\, \lambda_{\mu} \lambda_{\nu}],\\
R_{,\mu}&=&(A-\epsilon)\, \lambda_{\mu}+ \dot{z}_{\mu},
\end{eqnarray}

\noindent
where

\begin{equation}
A=\ddot{z}^{\mu}\,(x_{\mu}-z_{\mu}),~~~ \dot{z}^{\mu}\, \dot{z}_{\mu}=
\epsilon= 0, \pm 1.
\end{equation}

\noindent
Here $\epsilon =0,-1$ for null and time-like curves, respectively.
Furthermore, we have

\begin{equation}
\lambda_{\mu}\, \dot{z}^{\mu}=1,~~~ \lambda^{\mu}\, R_{, \mu}=1.
\end{equation}

\noindent
Letting $ a={A \over R}$, it is easy to prove that

\begin{equation}
a_{,\mu}\, \lambda^{\mu}=0. \label{cond}
\end{equation}

\noindent Similarly, other scalars $(a_{1}, a_{2}, \cdots)$,
satisfying the same property (\ref{cond}) obeyed by $a$ can be
defined

\begin{equation}
a_{k}\equiv \lambda_{\mu}\, {d^k \, \ddot{z}^{\mu} \over
d\tau_{0}^k}, ~~~k=1,2,\cdots ,n. \label{ak}
\end{equation}

\noindent
Moreover one has

\begin{equation}
a_{k, \alpha}\, \lambda^{\alpha}=0, \label{con01}
\end{equation}

\noindent for all $k$ ($k=0$ is also included if we let
$a_{0}=a$). For a more detailed discussion, please refer to
\cite{gu1}. Here $n$ is a positive integer which depends on the
dimension $D$ of the manifold ${\bf M_{D}}$. An analysis using
Serret-Frenet frames shows that the scalars $(a,\,a_{k})$, are
related to the curvature scalars of the curve $C$ in ${\bf
M_{D}}$. The number of such scalars is $D-1$ \cite{spv}. Hence we
let $n=D-1$.

\section{Massless Scalar Field }

\noindent
 Let $\phi$ describe a massless
scalar field satisfying the free field equation

\begin{equation}
\eta^{\mu \nu}\, {\partial^2 \, \phi \over \partial
x^{\mu} \partial x^{\nu}}=0. \label{scal}
\end{equation}

\noindent
Let $D$ be a positive even integer, and
 $\phi^{(D)}$ and $\phi^{(D+2)}$ denote
 the retarded solutions (Li{\' e}nard-Wiechert
potentials) of the massless scalar field in $D$ and $D+2$ dimensions,
respectively. Then

\begin{equation}
\phi^{(D+2)}={1 \over R} \, { d \over d\tau}\, \phi^{(D)}. \label{rec0}
\end{equation}

\noindent
In this recursion relation we emphasize that the expressions on the
right hand side are those of $D$-dimensions. Take the solution $\phi^{(D)}$ in
$D$-dimensions, take its $\tau$ derivative and divide this
by the $R$ of $D$-dimensions.
The  result is the solution $\phi^{(D+2)}$ of $D+2$-dimensions.
For the proof of  relation (\ref{rec0}) see Appendix B.
Below we explicitly give these solutions for $D=4,6,8,10$:

\begin{eqnarray}
&&\phi^{(4)}={c \over R},\\
&&\phi^{(6)}={1 \over R^2}\,[\dot{c}-pc], \label{fi6}\\
&&\phi^{(8)}={1 \over R^3}\,[\ddot{c}-3p \dot{c}+(-a_{1}+3p^2)c], \\
&&\phi^{(10)}={1 \over R^4}\,[{d^3 c \over d \tau^3}-6p \ddot{c}+(15p^2-
4a_{1})\dot{c} \nonumber \\
&&+(-a_{2}+10 p a_{1}-15p^3+{1 \over R}\, \dot{z}_{\alpha}
{d^3 z^{\alpha} \over d\tau^3})c],
\end{eqnarray}

\vspace{0.3cm}

\noindent where $c=c(\tau)$ is the (time dependent) scalar charge
and $p \equiv a-{\epsilon \over R}$.

 The flux of massless scalar field energy is then given by (see
 \cite{brt} and \cite{syng} for this definition and also for the integration surface
 $S$)

\begin{equation}
dE=-\int_{S}\, \dot{z}_{\mu}\,T_{\phi}^{\mu \nu} dS_{\nu}
\end{equation}

\noindent
where $T^{\phi}_{\mu \nu}=\partial_{\mu}\,\phi\, \partial_{\nu}\,\phi-
{1 \over 4} (\eta^{\alpha  \beta}
\,\partial_{\alpha}\,\phi\, \partial_{\beta}\, \phi) \eta_{\mu \nu}$
is the energy momentum tensor of the massless scalar field $\phi$.
The surface element $dS_{\mu}$ on $S$ is given by

\begin{equation}
dS_{\mu}=n_{\mu} R^{D-3} d\tau\, d\Omega,
\end{equation}

\noindent
where  $n_{\nu}$ is orthogonal to the velocity vector field
$\dot{z}_{\mu}$ which is defined through

\begin{equation}
\lambda_{\mu}=\epsilon \dot{z}_{\mu}+\epsilon_{1}\,{n_{\mu} \over R},
~~~ n^{\mu}\, n_{\mu}=-\epsilon R^2.
\end{equation}

\noindent
Here $\epsilon_{1}=\pm 1$.
For the remaining part of this work we shall assume $\epsilon=-1$
($C$ is a time-like curve).
One can consider $S$ in the rest frame as a sphere of radius $R$. Here
$d\Omega$ is the solid angle. Letting $dE/d\tau=N_{\phi}$, we have

\begin{equation}
N_{\phi}^{(D)}=-\int_{S^{D-2}}\, \dot{z}_{\mu}\,T_{\phi}^{\mu
\nu}\, n_{\nu} \, R^{D-3}\, d\Omega,
\end{equation}

\noindent 
where $S^{D-2}$ is the $(D-2)$-dimensional sphere centered
at $\tau=\tau_{0}$ on the curve $C$. At very large values of $R$
the energy flux is given by

\[
N_{\phi}^{(D)}=-\int_{S^{D-2}}\, d\Omega\, R^{D-3}\,(\dot{z}^{\alpha}\, \partial_{\alpha}\,
\phi)(n^{\beta}\, \partial_{\beta}\, \phi).
\]

\noindent
It turns out that the energy flux expression has a fixed sign for all $D$.
The energy flux of the massless scalar field $\phi$
as $R \rightarrow \infty$ is given by

\[
N_{\phi}^{(D)}=-\epsilon_{1} \int_{S^{D-2}}\, [\xi^{(D)}]^2\, d\Omega
\]

\noindent 
where we obtain $R$ independent functions (for each $D$) $\xi^{(D)}$ from

\[
\xi^{(D)}=\lim_{R \rightarrow \infty}\, \left [ R^{D \over 2}\, \phi^{(D+2)}
\right].
\]

\vspace{0.3cm}

\noindent 
As an example let $D=4$. We take $\phi^{(6)}$ from
(\ref{fi6}), multiply it by $R^2$ and let $R \rightarrow \infty$
(then $p \rightarrow a$), and finally we obtain $\xi^{(4)}$. The explicit
expressions of $\xi^{(D)}$ are as follows:

\begin{eqnarray}
\xi^{(4)}&=&\dot{c}-ac,\\
\xi^{(6)}&=&\ddot{c}-3a \dot{c}+(-a_{1}+3a^2)c, \\
\xi^{(8)}&=&{d^3 c \over d\tau^3} -6a \ddot{c}+(15a^2-4a_{1}) \dot{c}+
(-a_{2}+10aa_{1}-15a^3) c, \\
\xi^{(10)}&=&{d^4 c \over d\tau^4}-10a {d^3 c \over d\tau^3}+
(45 a^2-10 a_{1}) \ddot{c}-(5 a_{2}-60a a_{1}+105 a^3) \dot{c} \nonumber \\
&& -(a_{3}-15 a a_{2}-10 a_{1}^2 +105 a_{1} a^2-105 a^4)\,c.
\end{eqnarray}

\noindent
Hence we have (assuming $c=$ constant)

\begin{eqnarray}
N_{\phi}^{(4)}&=&-\epsilon_{1}({4\pi \over 3})\,c^2\, \kappa_{1}^2 ,\\
N_{\phi}^{(6)}&=&-\epsilon_{1}({8 \pi^2 \over 105})\, c^2 [20 \kappa_{1}^4+
7\dot{\kappa}_{1}^2+7 \kappa_{1}^2 \kappa_{2}^2],\\
N_{\phi}^{(8)}&=&-\epsilon_{1} ({16 \pi^3 \over 10395}) c^2
\{ 99 [(\ddot{\kappa}_{1}-4\kappa_{1}^3-\kappa_{1} \kappa_{2}^2)^2+
(2 \dot{\kappa}_{1} \kappa_{2}+\kappa_{1} \dot{\kappa}_{2})^2 \nonumber\\
&& +\kappa_{1}^2 \kappa_{2}^2 \kappa_{3}^2]+\kappa_{1}^2\, [
900 \kappa_{1}^4+1100 \kappa_{1}^2 \kappa_{2}^2+3597 \dot{\kappa}_{1}^2] \}.
\end{eqnarray}

\section{Electromagnetic Field}

In the Lorentz gauge ($\partial_{\mu}\, A^{\mu}=0$), the Maxwell equations
reduce to the wave equation for the vector potential $A_{\mu}$,
 $ \eta^{\mu \nu}\, \partial_{\mu}\, \partial_{\nu}\, A_{\alpha}=0$.
By using the  curve $C$,  we can construct divergence
free (Lorentz gauge) vector fields $A_{\alpha}$ satisfying the
wave equation outside the curve $C$ in any even dimension $D$.
Similar to the case of the massless scalar field,  such vectors obey
 the following recursion relation

\begin{equation}
A_{\mu}^{(D+2)}={1 \over R}\, {d \over d\tau}\, A_{\mu}^{(D)}. \label{rec2}
\end{equation}

\noindent
In the recursion relation above $A_{\mu}^{(D)}$ is the electromagnetic vector
potential in even $D$-dimensions, with $\mu = 0,1, \cdots , D-1$. On the
right hand side of the  recursion relation all operations are done in
$D$-dimensions, just like the scalar case. However the result is to be
considered as the electromagnetic vector potential of $D+2$-dimensions, with
$\mu=0,1, \cdots , D+1$ on the left hand side.
As an example we have $A_{\mu}^{(4)}=
{\dot{z}_{\mu} \over R}$ as the electromagnetic vector potential of four
dimensions \cite{jack}. Here $\dot{z}_{\mu}$ is the four velocity, $R$ and $\tau$
are, respectively, the retarded distance  and time in four dimensions. Using
the recursion relation (\ref{rec2}) the right hand side becomes
\[
{\ddot{z}_{\mu}-a \dot{z}_{\mu} \over R^2}+\epsilon {\dot{z}_{\mu} \over R^3}.
\]
We then regard this expression as the solution $A_{\mu}^{(6)}$
of the Maxwell field equations in
$6$-dimensions. Indeed it satisfies both the Lorentz condition and the field
equations of $6$-dimensions, as can be verified separately. Starting from
$D=4$, we can generate all even dimensional vector
potentials satisfying the Maxwell equations. For instance,
the vector potentials for $D=4,6,8,10$ are explicitly given by

\begin{eqnarray}
A_{\mu}^{(4)}&=&{\dot{z}_{\mu} \over R},\\
A_{\mu}^{(6)}&=&{1 \over R^2} [ \ddot{z}_{\mu}-p \dot{z}_{\mu}],\\
A_{\mu}^{(8)}&=&{1 \over R^3} [{d^3 z_{\mu} \over d\tau^3}\, -
3 p \ddot{z}_{\mu}+(-a_{1}+3 p^2) \dot{z}_{\mu}],\\
A_{\mu}^{(10)}&=&{1 \over R^4} [{d^4 z_{\mu} \over d\tau^4}
-6p {d^3 z_{\mu} \over d\tau^3} +(15p^2-4a_{1}) \ddot{z}_{\mu} \nonumber \\
&&+(-a_{2}+10pa_{1}-15p^3+
{1 \over R} \dot{z}_{\alpha}\, {d^3 z^{\alpha} \over d\tau^3})\, \dot{z}_{\mu}].
\end{eqnarray}

\noindent The flux of electromagnetic energy is then given by
\cite{brt} (the integration surface $S$ is also given in this
reference)

\begin{equation}
dE=-\int_{S}\, \dot{z}_{\mu}\,T_{e}^{\mu \nu} dS_{\nu},
\end{equation}

\noindent
where $T^{e}_{\mu \nu}=F_{\mu \alpha}\, F_{\nu}\,^{\alpha}-{1 \over 4} F^2
\eta_{\mu \nu}$ is the Maxwell energy momentum tensor, $F_{\mu \nu}=
A_{\nu,\mu}-A_{\mu,\nu}$ is the  electromagnetic field tensor
 and $F^2 \equiv F^{\alpha \beta}\,F_{\alpha \beta}$.

Letting $dE/d\tau=N_{e}$  \cite{syng2}, we have

\begin{equation}
N_{e}^{(D)}=-\int_{S^{D-2}}\, \dot{z}_{\mu}\,T_{e}^{\mu \nu}\, n_{\nu} \, R^{D-3}\,
 d\Omega.
\end{equation}

\noindent
At very large values of $R$, for all even $D$, we get

\begin{equation}
N_{e}^{(D)}=-\epsilon_{1} \int_{S^{D-2}}\, \xi^{(D)}_{\mu}\, \xi^{(D)}_{\nu}\,
\eta^{\mu \nu} d\Omega,
\label{los1}
\end{equation}

\noindent
where

\begin{equation}
\xi_{\mu}^{(D)}=\lim_{R \rightarrow \infty} [A_{\mu}^{(D+2)} \, R^{D \over 2}],
\label{gaug}
\end{equation}

\noindent
so that $\lambda^{\mu}\, \xi^{(D)}_{\mu}=0$ for all $D$.

Here we have two remarks. First one is on the gauge dependence of
(\ref{gaug}). The only gauge freedom  left in
our solutions is $A_{\mu} \rightarrow A_{\mu}+\partial _{\mu} \phi$,
where $\phi$ satisfies the scalar wave equation (\ref{scal}). However
we have already found the
solutions of the scalar wave equation for all even dimensions. It can be shown
that the contribution of such scalar functions to the norm of
$\xi_{\mu}^{(D)}$
is zero in the limit $R \rightarrow \infty$.
 Our second remark is on the sign of $N^{(D)}_{e}$ in (\ref{los1}).
The vectors $\xi^{(D)}_{\mu}$ in all even
dimensions are orthogonal to the null vector
$\lambda_{\mu}$, hence they must be  either i)\, space-like vectors,
ii)\, proportional to $\lambda_{\mu}$, or iii)\, zero vectors \cite{syng}.
They are zero only when
the curve $C$ is a straight line which leads to no radiation. They cannot be
proportional to the null vector $\lambda_{\mu}$ either,
because this again leads to the trivial case of
zero radiation. In the first three cases ($4$, $6$, $8$ dimensions)
it can be easily observed that
 zero radiation  implies that $\xi^{(D)}_{\mu}$ is a zero vector.
Hence $\xi^{(D)}_{\mu}$ is a space-like vector in all even
dimensions. Therefore the sign of the right hand side of (\ref{los1}) is the
same in all dimensions. These vectors are explicitly given as follows:

\begin{eqnarray}
\xi_{\mu}^{(4)}&=& \ddot{z}_{\mu}-a \dot{z}_{\mu},\\
\xi_{\mu}^{(6)}&=&{d^3 z_{\mu} \over d\tau^3}\, -3 a \ddot{z}_{\mu}+
(-a_{1}+3a^2) \dot{z}_{\mu},\\
\xi_{\mu}^{(8)}&=&{d^4 z_{\mu} \over d\tau^4}
-6a {d^3 z_{\mu} \over d\tau^3}\, +(15 a^2-4a_{1}) \ddot{z}_{\mu} \nonumber \\
&&+(-a_{2}+10aa_{1}-15a^3)\, \dot{z}_{\mu},\\
\xi_{\mu}^{(10)}&=&{d^5 z_{\mu} \over d\tau^5}-
10 a {d^4 z_{\mu} \over d\tau^4}+(45a^2-10 a_{1})
{d^3 z_{\mu} \over d\tau^3} \nonumber\\
&&+(-5 a_{2}+60 a a_{1}-105 a^3) \ddot{z}_{\mu} \nonumber \\
&&+(-a_{3}+15a a_{2}+10a_{1}^2-105 a_{1} a^2+105a^4) \dot{z}_{\mu}.
\end{eqnarray}

\noindent
These lead to the following energy flux expressions

\begin{eqnarray}
N_{e}^{(4)}&=&- \epsilon_{1}\,{8\pi \over 3} \, \kappa_{1}^2,\\
N_{e}^{(6)}&=&- \epsilon_{1}\,{32 \pi^2 \over 15} \,(\dot{\kappa}_{1}^2+
\kappa_{1}^2 \kappa_{2}^2+{9 \over 7} \kappa_{1}^4),\\
N_{e}^{(8)}&=&- \epsilon_{1}\,{32 \pi^3 \over 10395} \{297 [(\ddot{\kappa}_{1}
-{4 \over 3}\, \kappa_{1}^3-\kappa_{1}\, \kappa_{2}^2)^2+
(2\dot{\kappa}_{1}\, \kappa_{2}+\kappa_{1}\, \dot{\kappa}_{2})^2
\nonumber\\
&&+\kappa_{1}^2\, \kappa_{2}^2\, \kappa_{3}^2]
+4\kappa_{1}^2\, [300 \kappa_{1}^4+506 \kappa_{1}^2 \,
\kappa_{2}^2+825 \dot{\kappa}_{1}^2]\}.
\end{eqnarray}

\vspace{0.3cm}

\noindent
To be compatible with the classical results \cite{brt},
\cite{jack}, one should take $\epsilon_{1}= -1$.

\section{Conclusion}

In this work we have considered radiation of scalar and vector fields
due to  acceleration of point charged particles. We first examined the geometric properties
of their paths in an even dimensional Minkowski space ${\bf M_{D}}$.
By using the curve kinematics we developed,
we have first found the retarded solutions of the scalar field equations in
${\bf M_{D}}$. These solutions describe
the potentials of the accelerated scalar charges and
we have examined the energy loss due to such a radiation. We have shown
that in all even dimensions such scalar point particles lose energy.
We have given explicit examples for $D=4,6,8,10$. We then found the retarded
solutions of the Maxwell field equations that describe
the point particles carrying electric charges. Again,
using the curve kinematics we  developed
an algorithm to calculate the vector potential $A_{\mu}$ in $D+2$-dimensions
from the one in $D$-dimensions. We have given explicit examples for
$D=4,6,8$. We have calculated the energy flux in each case, and  we have shown
that particles lose energy due to acceleration in all even dimensions.

\vspace{1.5cm}

This work is partially supported by the Scientific and Technical Research
Council of Turkey and by the Turkish Academy of Sciences.

\section*{Appendix A: Serret-Frenet Frames.}

In this Appendix, we first give the Serret-Frenet frame in $D$ dimensions.
Here we shall assume that the
curve $C$ described in Section 2 is time-like and has the tangent vector
 $T^{\mu}=\dot{z}^{\mu}$.
Starting from this unit tangent vector, by repeated differentiation
with respect to
the arclength parameter $\tau_{0}$, one can generate an orthonormal frame
$\{ T^{\mu}, N_{1}^{\mu},N_{2}^{\mu}, \cdots ,N_{D-1}^{\mu} \}$ ,the
{\it Serret-Frenet frame}:

\begin{eqnarray}
\dot{T}^{\mu}&=&\kappa_{1}\, N_{1}^{\mu},  \\
\dot{N}_{1}^{\mu}&=&\kappa_{1}\,T^{\mu}-\kappa_{2}\, N_{2}^{\mu},  \\
\dot{N}_{2}^{\mu}&=&\kappa_{2}\,N_{1}^{\mu}-\kappa_{3}\,N_{3}^{\mu},\\
&& \cdots \nonumber\\
\dot{N}_{D-2}^{\mu}&=&\kappa_{D-2}\, N_{D-3}^{\mu}-
\kappa_{D-1}\, N_{D-1}^{\mu},\\
\dot{N}_{D-1}^{\mu}&=&\kappa_{D-1}\,N_{D-2}^{\mu}.
\end{eqnarray}

\noindent
Here $\kappa_{i},~ (i=1,2, \cdots , D-1)$ are the curvatures of the curve
$C$ at the point
$z^{\mu}(\tau_{0})$. The normal vectors $N_{i}, ~(i=1,2, \cdots , D-1)$ are
space-like unit vectors.
Hence at the point $z^{\mu}(\tau_{0})$ on the curve we have an orthonormal
 frame which
can be used  as a basis of the tangent space (of ${\bf M_{D}}$) at this point.
In Section 2,
we have defined some scalars
$$a_{k}={d^k \ddot{z}_{\mu} \over d\tau_{0}^k}\, \lambda^{\mu},$$
where
$$\lambda^{\mu}=\epsilon T^{\mu}+\epsilon_{1}\,
{n^{\mu} \over R}.$$

Here $n^{\mu}$ is a space-like
vector orthogonal to $T^{\mu}$. It can be expressed as a linear combination
of the unit vectors $N_{i}$'s as
$$n^{\mu}=\alpha_{1}\, N_{1}^{\mu}+\alpha_{2} \, N_{2}^{\mu}\, +
\cdots + \alpha_{D-1}\, N_{D-1}^{\mu},$$
where $\alpha_{1}^2+\alpha_{2}^2+\cdots +\alpha_{D-1}^2=R^2$.
One can choose the spherical angles $\theta , \phi_{1}, \cdots, \phi_{D-4}
\in (0, \pi),~~\phi_{D-3} \in (0, 2\pi) $ such that

\begin{eqnarray}
\alpha_{1}&=&R \cos \theta,~~ \alpha_{2}=R \sin \theta \cos \phi_{1},
~~\alpha_{3}=
R \sin \theta \sin \phi_{1} \, \cos \phi_{2}, \cdots,\nonumber \\
\alpha_{D-2}&=&R \sin \theta\, \sin \phi_{1} \cdots \sin \phi_{D-4}\,
\cos \phi_{D-3}, \nonumber \\
\alpha_{D-1}&=&R \sin \theta\, \sin \phi_{1} \cdots \sin \phi_{D-4}\,
\sin \phi_{D-3}. \nonumber
\end{eqnarray}

\noindent
Hence we can calculate the scalars $a_{k}$ in terms of the curvatures of the
curve $C$ and
the angles $(\theta, \phi_{1}, \cdots, \phi_{D-3})$ at the point
$z^{\mu}(\tau_{0})$. We need these
expressions in the evaluation of energy flux formulae. As an example we give
$a$ and $a_{1}$:

\begin{equation}
a=-\epsilon \epsilon_{1} \kappa_{1}\, \cos \theta,
~~a_{1}=\kappa_{1}^2-\epsilon \epsilon_{1} \dot{\kappa}_{1}
 \cos \theta+
\epsilon \epsilon_{1} \kappa_{1} \, \kappa_{2} \, \sin \theta \, \cos \phi_{1}.
\end{equation}

\noindent
The rest of the scalars can be determined similarly.
It is clear that these scalars, $a_{k}$,
depend on the curvatures and the spherical angles, for all $k$.

\section*{Appendix B: The proof of the Recursion relations (11) and (27)}

Here we give the proof for the vector potential case. The same type of
proof applies also for the scalar case.
Using the recursion relation (\ref{rec2}) successively we get

\begin{equation}
A_{\mu}^{(D)}=\left({1 \over R}\, {d \over d\tau}\right)^{{D \over 2}-2}\,\,
{\dot{z}_{\mu} \over R}. \label{rec3}
\end{equation}

\noindent
On the other hand, from \cite{gal} and \cite{kaz}, we have

\begin{equation}
A_{\mu}^{(D)}=\int \, G(x-z(\tau))\, \dot{z}_{\mu}\, d\tau,
\label{vec1}
\end{equation}

\noindent where $\tau$ is the parameter of the curve $C$. The
integral here is carried on the range of $\tau \in (-\infty, \infty)$.
Here $G(x-z(\tau))$ is the retarded Green function given by

\begin{equation}
G(x-z(\tau))=\theta(x^{0}-z^{0}) \, \delta^{{D \over 2}-2}(\Phi).
\label{grn}
\end{equation}

\noindent Here $\Phi$ is the world function given by
(\ref{dist}), $\theta(x)$ is the Heaviside step function and
$\delta^{k}(x) \equiv {d^k \over dx^k}\, \delta(x)$. Here we assume that
$D$ is an even integer (When $D$ is an odd integer, the expression
for the Green function in (\ref{grn}) contains the step function
instead of the $\delta$-function. Hence the potentials in all odd
dimensions remain non-local (integral expressions). This makes our
curve kinematics ineffective). The zeros of $\Phi$  denote the
advanced and retarded proper times on the curve $C$, but the step
function $\theta(x^{0})$ chooses the retarded one. Since the
integration is over the curve parameter $\tau$ in (\ref{vec1}), it
is  better to transform the derivative of
 the delta function with respect to $\Phi$ to the derivative with respect
to $\tau$. As a simple example consider the D=6 case

\begin{equation}
{d \over d\Phi}\, \delta (\Phi)= \left[ {1 \over d\Phi/d\tau}\, {d
\over d\tau}\,\delta (\Phi) \right]_{\Phi=0}. \label{delder}
\end{equation}

\noindent It is easy to show that ${d\Phi \over d\tau}=-R$. The
delta function $\delta(\Phi)$ can be expressed as follows

\[
\delta(\Phi)={\delta(\tau-\tau_{0}) \over R}
+{\delta(\tau-\tau_{1}) \over R }.
\]

\noindent
The second term will vanish identically due to the step function
in (\ref{grn}). Hence

\[
A^{(6)}_{\mu}={1 \over R} {d \over d\tau}\, {\dot{z}_{\mu} \over R},
\]

\noindent or simply $A^{(6)}={1 \over R}\, {d \over d\tau}\,
A^{(4)}$. This verifies our relation (\ref{rec2}). For the general
case, we need higher order derivatives of $\delta(\Phi)$ at
$\Phi=0$. We find such terms by using (\ref{delder}) and taking
successive derivatives. In the general case, for all $k=0,1,2,
\cdots $ we obtain (when $\Phi=0$)

\begin{equation}
{d^{k} \over d\Phi^{k}}\, \delta (\Phi)= \left[\left({-1 \over
R}\,{d \over d\tau} \right)^{k}\,\delta (\Phi) \right].
\end{equation}

\noindent Using this expression in the Green's function
(\ref{grn}) for $k={D \over 2}-2$, inserting it in the integral
equation (\ref{vec1})

\begin{eqnarray}
A_{\mu}^{(D)}&=&\int\,
\theta(x^{0}-z^{0})\,\delta^{{D \over 2}-2}(\Phi)\,\dot{z}_{\mu}\, d\tau,\\
&=& \int\, \theta(x^{0}-z^{0})\, \left({-1 \over R}\,{d \over
d\tau} \right)^{{D \over 2}-2}\,\delta (\Phi)\, \dot{z}_{\mu}\,
d\tau,
\end{eqnarray}

\noindent and integrating by parts, we obtain (\ref{rec3}).


\end{document}